\documentclass{article}
\usepackage{graphicx}
\graphicspath{%
    {converted_graphics/}% inserted by PCTeX
    {/}% inserted by PCTeX
}
\begin{document}

\begin{center}
{\LARGE Observations of an Edge-enhancing Instability}\vskip6pt

{\LARGE in Snow Crystal Growth near -15 C}\vskip6pt

{\Large Kenneth G. Libbrecht}\vskip4pt

{\large Department of Physics, California Institute of Technology}\vskip-1pt

{\large Pasadena, California 91125}\vskip-1pt

\vskip18pt

\hrule\vskip1pt \hrule\vskip14pt
\end{center}

\textbf{Abstract. }We present observations of the formation of plate-like
snow crystals that provide evidence for an edge-enhancing crystal growth
instability. This instability arises when the condensation coefficient
describing the growth of an ice prism facet increases as the width of the
facet becomes narrower. Coupled with the effects of particle diffusion, this
phenomenon causes thin plate-like crystals to develop from thicker prisms,
sharpening the edges of the plates to micron or sub-micron dimensions as
they grow. This sharpening effect is largely responsible for the formation
of thin plate-like ice crystals from water vapor near -15 C, which is a
dominant feature in the snow crystal morphology diagram. Other faceted
crystalline materials may exhibit similar morphological growth instabilities
that promote the diffusion-limited growth of plate-like or needle-like
structures.

\section{Introduction}

The formation of complex structures during solidification often results from
a subtle interplay of non-equilibrium, nonlinear processes, for which
seemingly small changes in molecular dynamics at the nanoscale can produce
large morphological changes at all scales. One popular example of this
phenomenon is the formation of snow crystals, which are ice crystals that
grow from water vapor in an inert background gas. Although this is a
relatively simple physical system, snow crystals display a remarkable
variety of columnar and plate-like forms, and much of the phenomenology of
their growth remains poorly understood. (For a review of the physics of snow
crystal growth, see \cite{libbrechtreview}.)

Observations of snow crystal growth dating back to the 1930s \cite{nakaka, libbrechtreview} reveal a complex and puzzling
dependence on temperature. At pressures near one bar and water vapor supersaturation levels often found
in clouds, for example, ice crystals typically grow into thin plate-like
forms near -2 C, slender columns and needles near -5 C, very thin plates
again near -15 C, and columns again below -30 C. The observed variations in
snow crystal structure with temperature and supersaturation are often
displayed in a snow crystal morphology diagram, as shown in Figure \ref%
{morphdiagram}. Despite more than a half-century of study, we still do not
understand the basic physical mechanisms that are responsible for the
unusual temperature-dependent morphologies of growing ice crystals \cite%
{libbrechtreview, nelson, pruppacher}.

\begin{figure}[htb] % float placement: (h)ere, page (t)op, page (b)ottom, other (p)age
  \centering
  % file name: C:/1KGLaaa/aatempfold/VIGproject/Paper3-EdgeInstability/arXiv/SnowMorph.jpg
  \includegraphics[width=5in,height=3.64in,keepaspectratio]{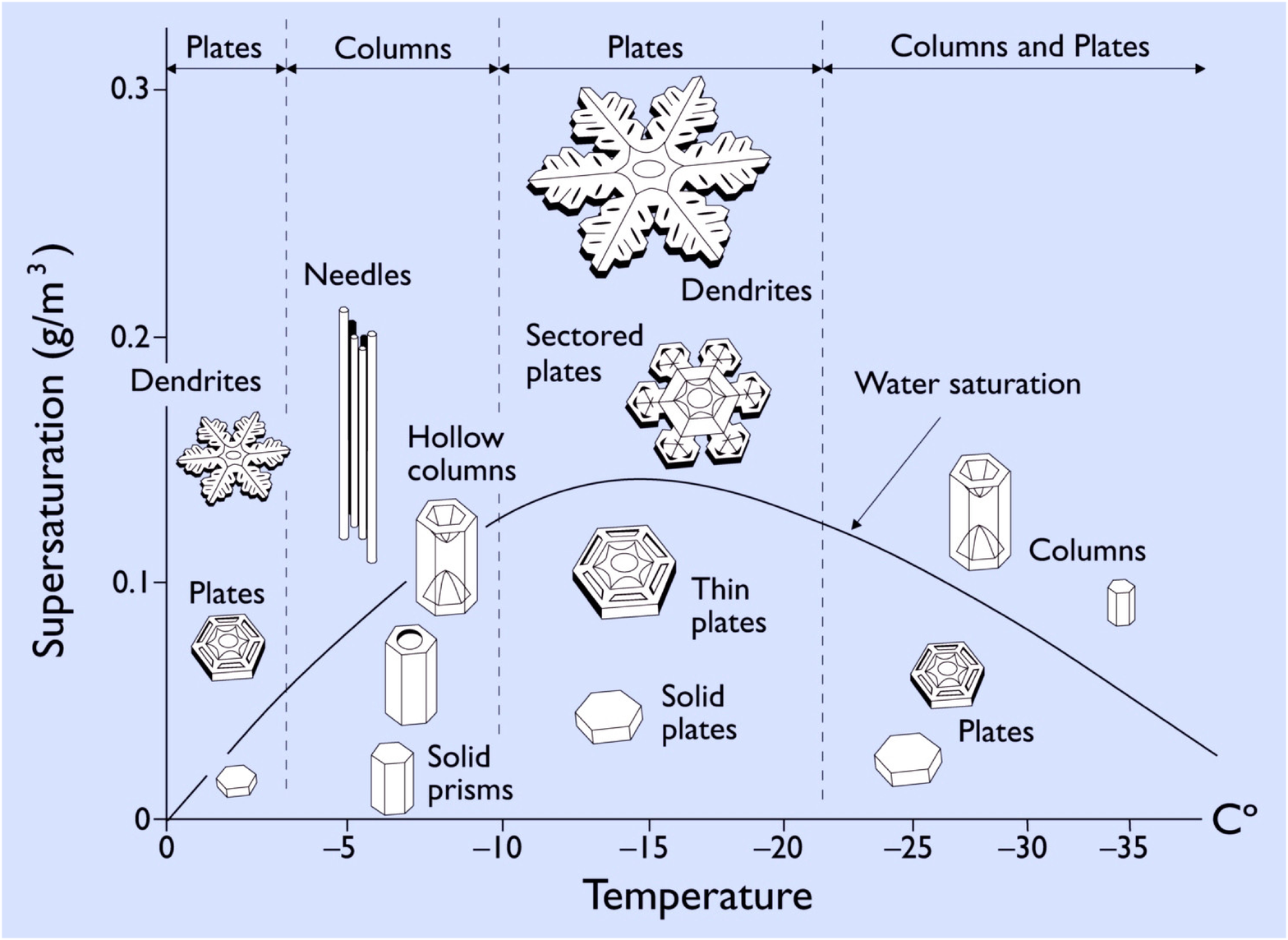}
  \caption{A snow crystal morphology
diagram showing the variations in growth forms as a function of temperature
and supersaturation for crystals grown in air at a pressure of one bar (from 
\protect\cite{libbrechtreview}). The water saturation line shows the supersaturation of
supercooled liquid water with respect to ice. Note that thin plate-like
crystals appear only in narrow temperature ranges near -2 C and -15 C.}
  \label{morphdiagram}
\end{figure}

Our focus here will be on the growth of thin plates near -15 C, as the
largest and thinnest crystals form in this temperature region, making this a
dominant feature in the morphology diagram. Aspect ratios of $%
L_{a}/L_{c}\approx 100$ or larger are commonly found with atmospheric ice
crystals, where $L_{a}$ is the crystal size along the a-axis of the crystal
(roughly equal to the diameter of the plate), and $L_{c}$ is the size of the
plate in the c-direction (equal to the plate thickness). Large stellar
plates typically have plate diameters of 2-4 mm with thicknesses of 10-20 $%
\mu $m \cite{nakaka}.

In modeling the growth of ice crystals from water vapor, we write the
perpendicular growth velocity of a surface in the usual Hertz-Knudsen form 
\begin{eqnarray}
v &=&\alpha \frac{\Omega (p_{surf}-p_{sat})}{\sqrt{2\pi mkT}}  \label{hertz}
\\
&=&\alpha v_{kin}\sigma _{surf}  \nonumber
\end{eqnarray}%
where $p_{surf}$ is the water vapor pressure at the growing surface, $p_{sat}
$ is the water vapor pressure in equilibrium with ice, $m$ is the mass of a
water molecule, and $\Omega $ is the molecular volume \cite{libbrechtreview}%
. This relation defines the condensation coefficient, $\alpha ,$ that
parameterizes the attachment kinetics at the ice surface, as well as the
kinetic velocity $v_{kin}$, where $\sigma _{surf}=(p_{surf}-p_{sat})/p_{sat}$
is the water vapor supersaturation at the surface. In general $\alpha $ will
be different for the two principal facets of ice, and for fast kinetics on
non-faceted surfaces we expect $\alpha \rightarrow 1.$

If the supersaturation had a constant value independent of location near the
ice surface, then the growth of a thin plate would imply $\alpha
_{prism}/\alpha _{basal}\approx v_{prism}/v_{basal}\approx L_{a}/L_{c}\gg 1.$
Diffusion modeling shows that the supersaturation at the edges of a thin
plate is not substantially higher than at the middle of the basal facets,
and for a large fast-growing plate the minimum supersaturation is actually
found at the edges of the plate \cite{libbrechtreview}. Thus
diffusion-limited growth will generally not promote the growth of thin
plates in the absence of a large $\alpha _{prism}/\alpha _{basal}$ ratio,
and in many cases diffusion requires that we must have $\alpha
_{prism}/\alpha _{basal}>L_{a}/L_{c},$ which further accentuates the
difference in attachment kinetics on the two facet surfaces. 

The problem with this simple inference is that measurements of ice crystal
growth rates for faceted surfaces have yielded values of $\alpha
_{prism}/\alpha _{basal}$ that are far smaller than what is needed to form
thin plate-like crystals. For example, numerous researchers have found that
ice crystals grow into roughly isometric shapes at low background pressures,
where effects from particle diffusion are reduced \cite{gonda, beckman, lamb}%
, suggesting that $\alpha _{prism}/\alpha _{basal}$ is near unity for the
conditions in these experiments. Since the growth of thin plates requires $%
\alpha _{prism}/\alpha _{basal}\gg 1,$ the discrepancy between low-pressure
and high-pressure measurements necessarily requires a nontrivial explanation.

In an earlier paper we proposed a mechanism to solve this problem that we
called \textit{structure-dependent attachment kinetics} (abbreviated \textit{%
SDAK}) \cite{sdak}. In essence, we proposed that $\alpha _{prism}$ depends
strongly on the structure of the prism facet surface, such that $\alpha
_{prism}$ increases when the width of the prism facet decreases, becoming
near unity when the width approaches atomic dimensions. For the case of a
thin plate, the width of the facet surface on the edge of the plate is
approximately $w\approx \sqrt{aR},$ where $a$ is the molecular step height
and $R$ is the effective radius of curvature of the edge of the growing
plate \cite{sdak}. Taking $R\approx 0.5$ $\mu $m for a typical thin plate
yields $w\approx 40a$ in the case of ice. We proposed that the SDAK
mechanism leads to an edge-enhancing growth instability that promotes the
formation of thin plates.

Below we present observations of a hysteresis behavior in the growth of thin
plates that strongly supports our proposed SDAK mechanism. Quantitative
measurements reveal that $\alpha _{prism}$ is indeed substantially larger on
the edge of a thin plate than on a flat prism facet. This behavior then
leads to an edge-enhancing growth instability that is responsible for the
formation thin plate-like crystals near -15 C.

\section{Observations}

\subsection{Intrinsic Growth Rates of Ice Facets}

We recently published improved measurements of the growth rates of faceted
basal surfaces as a function of temperature \cite{basalgrowth}, indicating $%
\alpha _{basal}\approx \exp (-\sigma _{basal}/\sigma _{surf})$ over a broad
temperature range, and specifically $\sigma _{basal}=\sigma
_{basal,0}\approx 2.0$ percent at -15 C. Similar measurements of prism
growth rates \cite{unpub} give the same functional form for $\alpha _{prism}$
at -15 C, and for flat prism facets we have $\sigma _{prism}=\sigma
_{prism,0}\approx 4.2$ percent. These measurements were made in a background
of air at a pressure of approximately 0.03 bar, since lower pressures reduce
effects from diffusion-limited growth. Similar measurements at 1 bar yield $%
\alpha $ values that are consistent with the low-pressure data \cite%
{basalgrowth}. These measurements are roughly consistent with previous data,
although several improvements in our experimental techniques have yielded $%
\sigma _{prism,0}$ values that are higher than quoted in \cite%
{libbrechtreview}.

These data indicate that $\alpha _{prism}/\alpha _{basal}\approx \exp
(-\Delta \sigma /\sigma )$ with $\Delta \sigma =2.2$ percent at -15 C. Since
this ratio is less than unity for all supersaturations, it suggests (in the
absence of the SDAK mechanism) that the intrinsic growth form at -15 C
should be a columnar crystal. Demonstrating this intrinsic columnar growth
directly in low-pressure observations is difficult, however, since any ice
surface contacting a substrate may be affected by substrate interactions.
Generally we have found that these interactions reduce the nucleation
barrier and yield untrustworthy growth measurements, especially at low
supersaturations when the intrinsic 2D nucleation rate is low. Nevertheless,
when one basal facet is flat against the substrate (as in \cite{basalgrowth}%
), we do occasionally observe the growth of tall columns. Since substrate
interactions cannot reduce the nucleation barrier all along a tall column,
our interpretation of these observations is that sometimes the substrate
interactions are greatly reduced, due to some impurity coating on the
substrate that we have not yet determined. When this happens we witness the
intrinsic growth rates on all facets, giving columnar crystals. More
typically substrate interactions increase the growth of the prism facets via
enhanced nucleation where the crystal contacts the substrate, resulting in
more isometric ice prisms.

Our overall conclusion from this series of measurements is that the
intrinsic attachment coefficients for large facet surfaces are adequately
described by the functional form above with $\sigma _{basal,0}\approx 2.0$
percent and $\sigma _{prism,0}\approx 4.2$ percent at -15 C. Growth at low
background pressures yields mainly large faceted surfaces, and the available
evidence suggests that the intrinsic growth form at -15 C may indeed be a
columnar crystal. Additional low-pressure measurements, preferably with free
falling or levitated crystals \cite{levitated} to avoid substrate
interactions, are needed to confirm this surprising conclusion. The fact
that thin plates are usually observed at -15 C is then entirely because of
the SDAK mechanism. The resulting edge-enhancing instability appears only at
higher background pressures, since the instability depends in part on the
effects of diffusion-limited growth. We now examine additional measurements
supporting this claim.

\subsection{Plate-on-Pedestal Growth}

We begin with a detailed description of the growth of a representative
crystal at -15 C in air at a pressure of one bar. Using the apparatus
described in \cite{vigapparatus}, we began the experiment by dropping a
small hexagonal plate crystal onto a temperature-controlled sapphire
substrate. The initial radius of this crystal (taken to be half the distance
from one prism facet to the opposite facet) was 13.5 $\mu $m, and the
initial thickness was 2.5 $\mu $m. The radius was determined by direct
optical microscopy while the thickness was measured by broad-band
interferometry, as described in \cite{vigapparatus}. Initially the
temperature $T_{IR}$ of an ice reservoir above the crystal was equal to the
substrate temperature $T_{subst}$, so the supersaturation was equal to zero.
We then slowly increased the ice reservoir temperature and thus the
supersaturation above the test crystal, causing it to grow.

\begin{figure}[p] % float placement: (h)ere, page (t)op, page (b)ottom, other (p)age
  \centering
  % file name: C:/1KGLaaa/aatempfold/VIGproject/Paper3-EdgeInstability/arXiv/TimeSeries.gif
  \includegraphics[bb=0 0 993 1237,width=5in,height=6.23in,keepaspectratio]{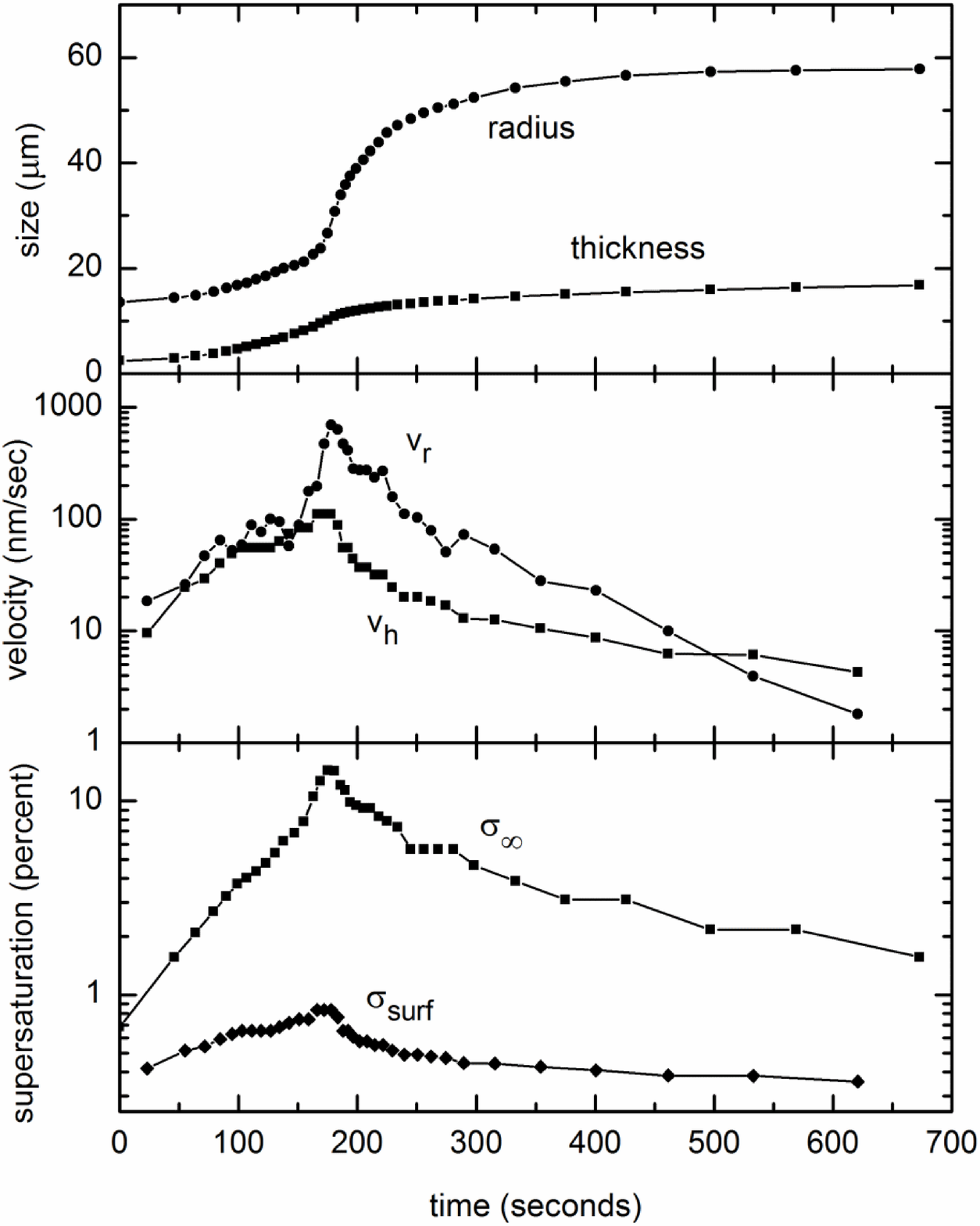}
  \caption{An example of \textquotedblleft
plate-on-pedestal\textquotedblright\ ice crystal growth. Top: The radius and
thickness of the test crystal as a function of time. One basal surface of
the crystal was resting on a substrate, and the test chamber was filled with
air at a pressure of one bar. Middle: The basal $v_{h}$ and prism $v_{r}$
growth velocities. Bottom: The supersaturation far above the crystal $%
\protect\sigma _{\infty }$ and the supersaturation at the basal surface $%
\protect\sigma _{surf}.$ The latter quantity was determined from the basal
growth velocity using Equation \protect\ref{hertz}, assuming the measured
condensation coefficient $\protect\alpha _{basal}=\exp (-\protect\sigma %
_{basal,0}/\protect\sigma _{surf})$ described in the text.}
  \label{timeseries}
\end{figure}

\begin{figure}[p] % float placement: (h)ere, page (t)op, page (b)ottom, other (p)age
  \centering
  % file name: C:/1KGLaaa/aatempfold/VIGproject/Paper3-EdgeInstability/arXiv/0528-1-composite1.jpg
  \includegraphics[width=5in,height=6.45in,keepaspectratio]{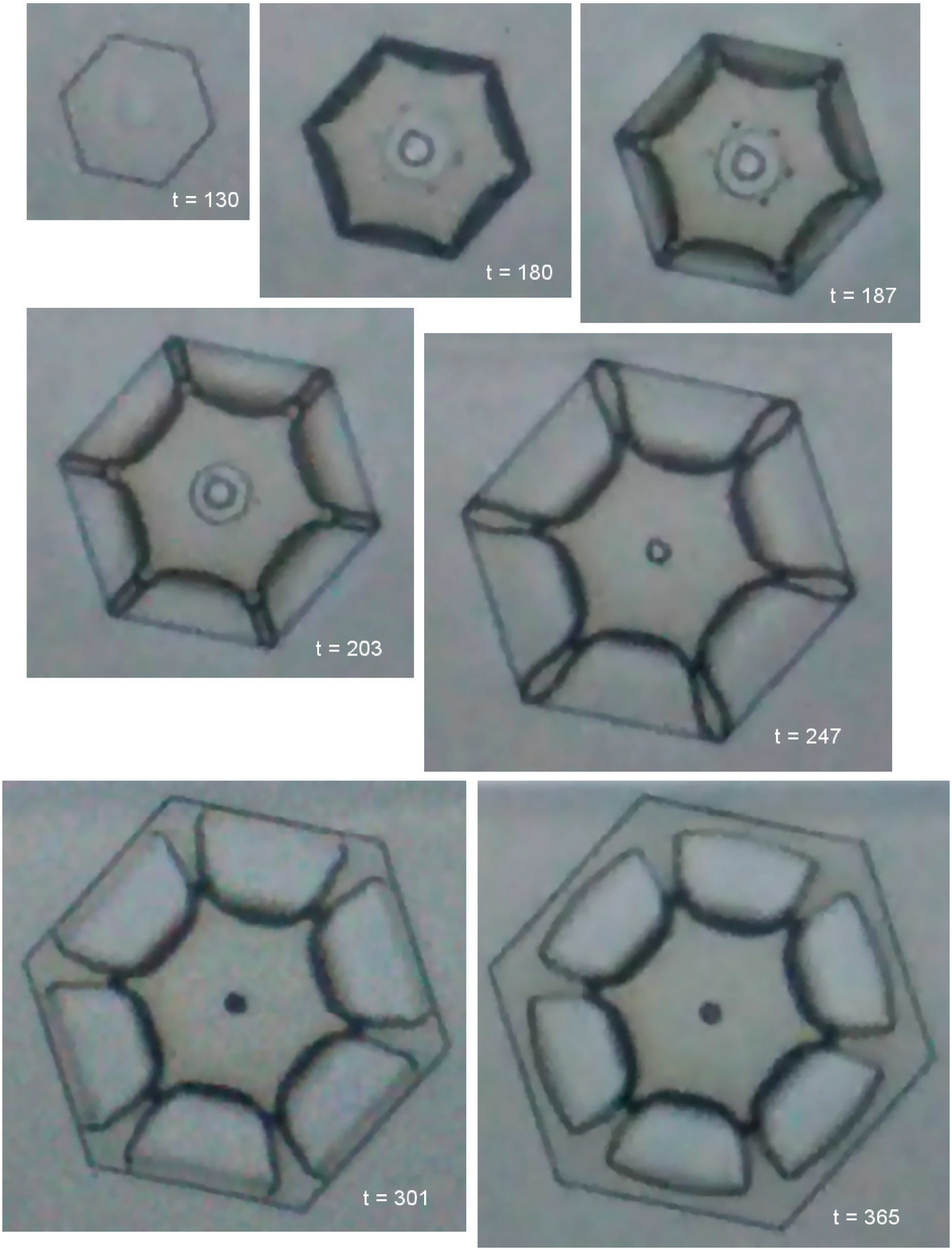}
  \caption{An example of \textquotedblleft
plate-on-pedestal\textquotedblright\ ice crystal growth. Each image is
labeled with the time in seconds, and the crystal sizes and growth
velocities at each time are given in Figure \protect\ref{timeseries}.
Between $t=130$ and $t=180$ a thin plate began growing from the top edge of
the ice prism. Throughout its subsequent growth this plate did not contact
the substrate except via its supporting pedestal. By $t=301$ the outer edge
of the plate had begun to thicken as the radial velocity diminished.}
  \label{composite}
\end{figure}

The crystal radius, its thickness, and the growth velocities as a function
of time are shown in Figure \ref{timeseries}, along with the supersaturation
above the crystal $\sigma _{\infty }$. Note that $\sigma _{\infty }$ was
changed with time during the experiment by changing $T_{IR},$ giving $\sigma
_{\infty }(t)$ as shown in the Figure, and this was the only external
experimental parameter that was changed during the run. Since the distance
from the substrate to the ice reservoir was much greater than the size of
the test crystal, $\sigma _{\infty }$ can be assumed (for modeling purposes)
to be the supersaturation at a hemispherical boundary far away from the
crystal.

As $\sigma _{\infty }$ was increased, the ice prism initially maintained its
simple shape, namely that of a thick hexagonal plate. At roughly $t\approx
160$ seconds, the morphology of the crystal changed as a thin plate-like
crystal began to grow out from the upper edge of the prism. We call this
\textquotedblleft plate-on-pedestal\textquotedblright\ growth because the
thin plate grew outward above the substrate, supported by a central
pedestal, like one half of a capped column crystal \cite{libbrechtreview}.

Figure \ref{composite} shows images of the plate-on-pedestal crystal at
various stages during its growth. The morphology of the thin plate during
its rapid growth phase is essentially that of a sectored plate crystal \cite%
{libbrechtreview}, and from additional observations we determined that the
ridges along the $a$-axes of the crystal were on the underside of the plate,
on the basal surface nearest the substrate. The top of the plate appeared to
be a flat basal facet surface without ridges. In Figure \ref{timeseries} the
radius refers to that of the top plate, while the thickness gives the
overall thickness of the crystal from the substrate to the top basal
surface. Once $\sigma _{\infty }$ was reduced, the edge of the thin plate
thickened, and eventually a thick \textquotedblleft rib\textquotedblright\
appeared around the circumference of the plate, as seen in the last image in
Figure \ref{composite}.

\begin{figure}[t] % float placement: (h)ere, page (t)op, page (b)ottom, other (p)age
  \centering
  % file name: C:/1KGLaaa/aatempfold/VIGproject/Paper3-EdgeInstability/arXiv/vhvrplot.gif
  \includegraphics[bb=0 0 1239 984,width=4.5in, keepaspectratio]{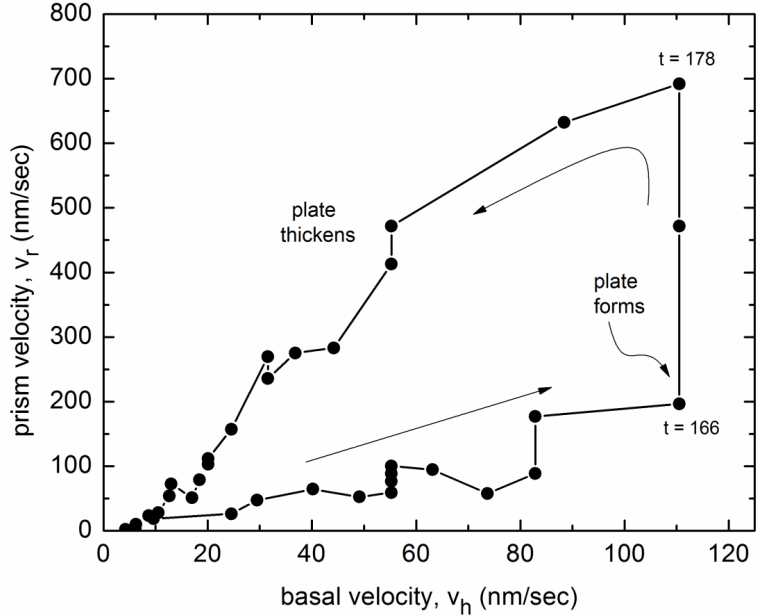}
  \caption{A plot of the basal growth
velocity $v_{h}$ and prism growth velocity $v_{r}$ of the test crystal (see
also Figure \protect\ref{timeseries}). By $t=166$ the thin plate had formed and
the prism velocity quickly increased. After $t=178$ the supersaturation was
reduced and the edge of the plate thickened as both $v_{h}$ and $v_{r}$
became smaller.}
  \label{vhvr}
\end{figure}

It is instructive to examine a $v_{h}$-$v_{r}$ plot of the crystal growth
shown in Figure \ref{vhvr}, as this exhibits a characteristic hysteresis
behavior. In the early stages of the experiment, both $v_{h}$ and $v_{r}$
increased slowly as $\sigma _{\infty }$ was increased, in this case with $%
v_{r}\approx 1.5v_{h}.$ Since the prism facets were in contact with the
substrate during this phase of the growth, $v_{r}$ was likely influenced by
substrate interactions that increased nucleation on the prism facets and
thus increased $v_{r}$ compared to that for a free-standing prism facet
surface. Had substrate interactions been absent, we would expect $v_{r}<v_{h}
$ because $\sigma _{prism,0}>\sigma _{basal,0}.$ By $t=166$ the thin plate
had begun to grow out from the top edge of the ice prism, which changed the
structure of the prism surface from a broad facet surface to a sharp edge.
According to our interpretation of the subsequent growth behavior, the sharp
edge of the prism surface resulted in a higher $\alpha _{prism}$ (and thus a
higher $v_{r})$ because of the SDAK effect. This brought into play a growth
instability that sharpened the edge of the plate further, leading to a rapid
increase in $v_{r}$ while $v_{h}$ (and thus the supersaturation near the
crystal surface) remained essentially constant. Around $t=178$ the
supersaturation $\sigma _{\infty }$ was reduced as shown in Figure \ref%
{timeseries}, which caused $v_{h}$ to drop. Because the edge of the plate
was still sharp, however, $\alpha _{prism}$ (and thus $v_{r}$) remained
high. As $\sigma _{\infty }$ was reduced further, eventually the edge of the
plate thickened and $\alpha _{prism}$ dropped back down as it approached the
intrinsic value for a flat facet.

\begin{figure}[t] % float placement: (h)ere, page (t)op, page (b)ottom, other (p)age
  \centering
  % file name: C:/1KGLaaa/aatempfold/VIGproject/Paper3-EdgeInstability/arXiv/SigmaCalcs.gif
  \includegraphics[bb=0 0 1232 970,width=4.5in, keepaspectratio]{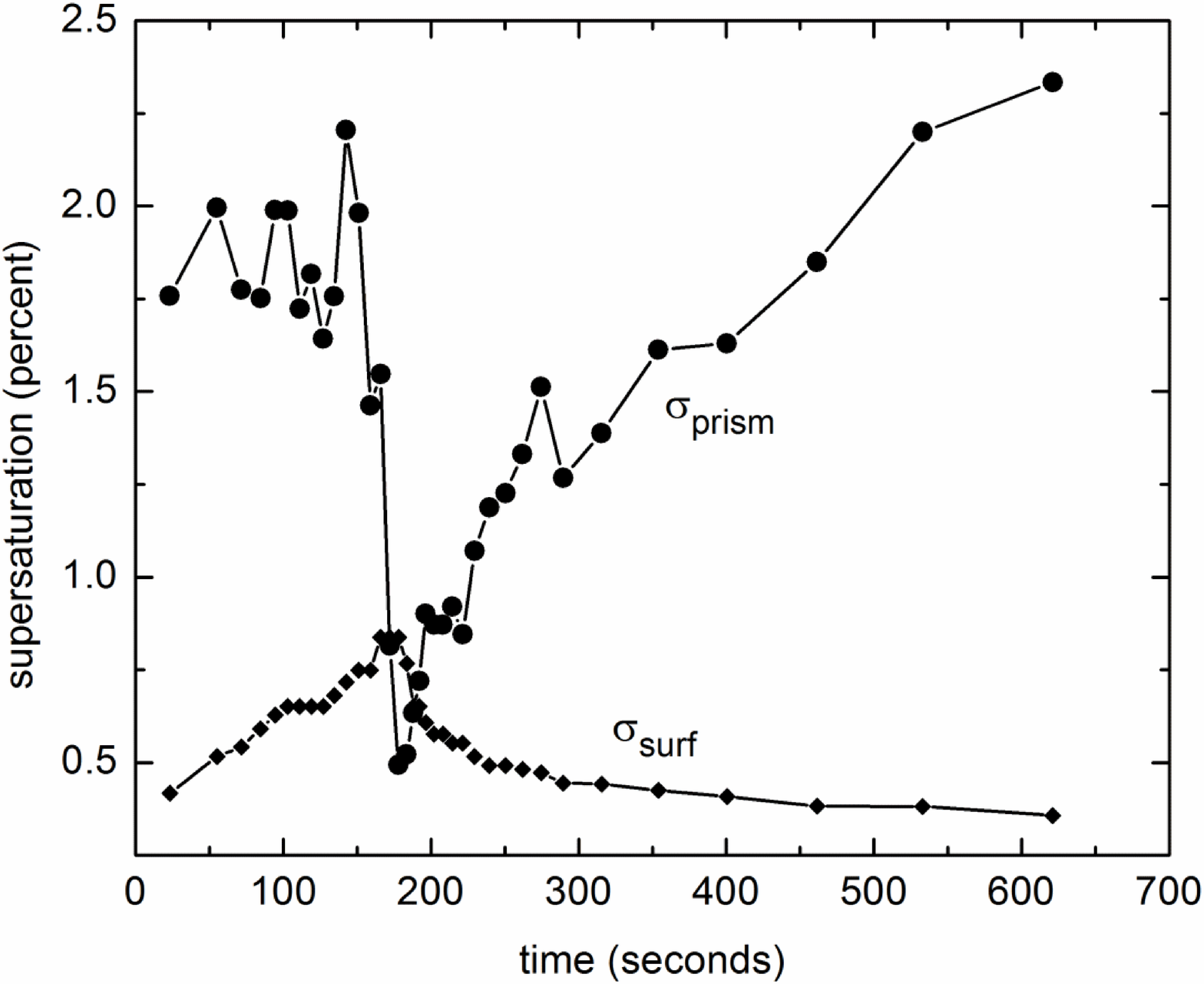}
  \caption{Estimates of $\protect\sigma %
_{surf}$ and $\protect\sigma _{prism}$ as the test crystal grew, using the
basal facet as a \textquotedblleft witness surface\textquotedblright\ as
described in the text.}
  \label{sigmacalcs}
\end{figure}

The supersaturation field around the growing crystal was determined by
diffusion, and we had no direct measure of $\sigma _{surf}.$ However we can
use the flat basal facet as a \textquotedblleft witness
surface\textquotedblright\ to estimate $\sigma _{surf}$ as follows. Since we
know $\alpha _{basal}(\sigma _{surf})=\exp (-\sigma _{basal,0}/\sigma
_{surf})$ from other measurements of the growth of large basal facets, we
can determine $\sigma _{surf}$ from this and the measured basal velocity $%
v_{h}$ using Equation \ref{hertz}. The result is plotted in the bottom panel
of Figure \ref{timeseries}. We then further assume that $\alpha
_{prism}=\exp (-\sigma _{prism}/\sigma _{surf})$ and use $v_{r}$ and
Equation \ref{hertz} to estimate $\sigma _{prism},$ with the result shown in
Figure \ref{sigmacalcs}. Although somewhat crude, this quantitative estimate
clearly confirms our hypothesis that the rapid increase in $v_{r}$ when the
thin plate formed is indicative of a rapid increase in $a_{prism}.$ Modeling
the attachment coefficient as $\alpha _{prism}=\exp (-\sigma _{prism}/\sigma
_{surf}),$ this indicates that $\sigma _{prism}\ll \sigma _{prism,0}$ on the
edge of the thin plate, as expected from the SDAK mechanism.

\begin{figure}[p] % float placement: (h)ere, page (t)op, page (b)ottom, other (p)age
  \centering
  % file name: C:/1KGLaaa/aatempfold/VIGproject/Paper3-EdgeInstability/arXiv/timeseries1.gif
  \includegraphics[bb=0 0 1075 1336,width=5.5in,height=6.84in,keepaspectratio]{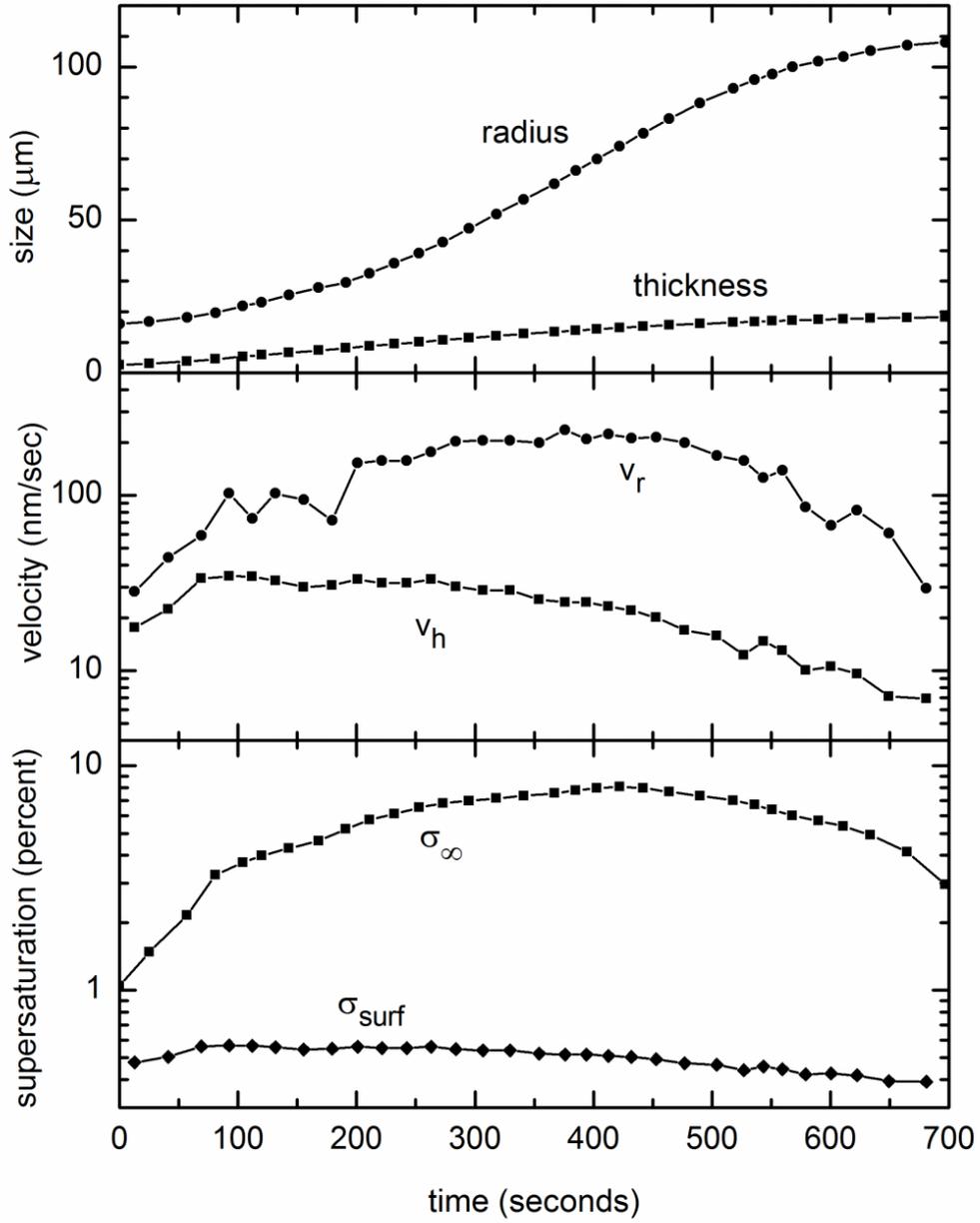}
  \caption{A second example crystal grown
at -15 C, showing a more gradual increase in $\protect\sigma _{\infty }$
with time during the early stages of growth.}
  \label{timeseries1}
\end{figure}

\begin{figure}[p] % float placement: (h)ere, page (t)op, page (b)ottom, other (p)age
  \centering
  % file name: C:/1KGLaaa/aatempfold/VIGproject/Paper3-EdgeInstability/arXiv/hysteresis2a.jpg
  \includegraphics[width=5in,height=7.25in,keepaspectratio]{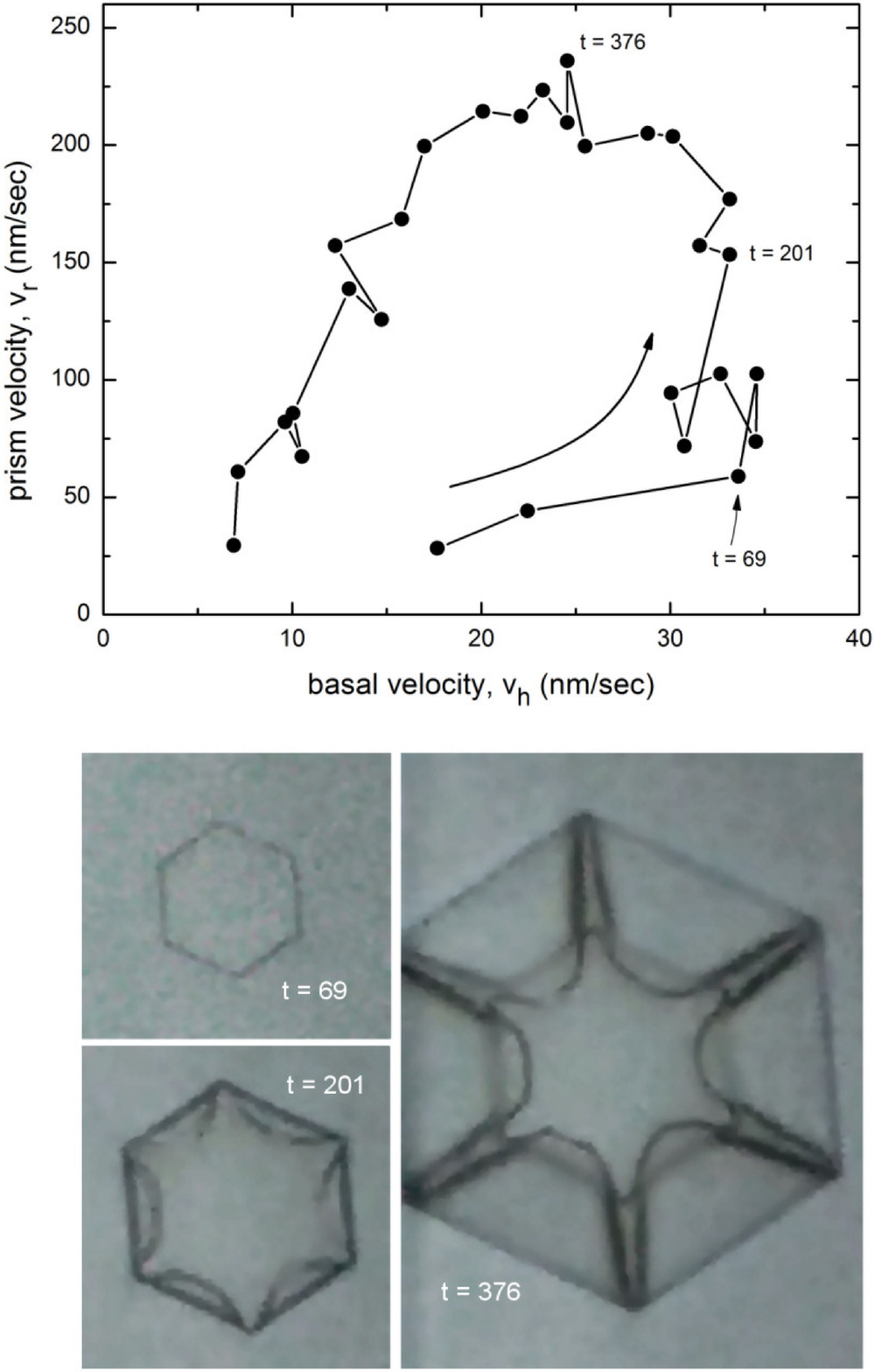}
  \caption{Top: A plot of the basal growth
velocity $v_{h}$ and prism growth velocity $v_{r}$ of the test crystal shown
in Figure \protect\ref{timeseries1}. Bottom:\ Three images of the crystal at
different times during its growth.}
  \label{hysteresis2}
\end{figure}

A key point in this analysis is an examination of what happened to this test
crystal around $t=166.$ What caused $v_{r}$ to suddenly increase while $%
v_{h} $ remained essentially constant, as seen in Figure \ref{vhvr}?
Essentially the only change that occurred was that a plate-like appendage
began growing on the top edge of the prism facet, as seen in Figure \ref%
{composite}. This alone would not substantially alter the diffusion field
around this small crystal, as is evidenced by the fact that $v_{h}$ remains
roughly unchanged. Since the diffusion field did not change abruptly near $%
t=166$, the rapid increase in $v_{r}$ means that $\alpha _{prism}$ increased
threefold while $\sigma _{surf}$ remained essentially constant. Our
conclusion, therefore, is that $\alpha _{prism}$ depends on the structure of
the prism facet, which confirms our original hypothesis of
structure-dependent attachment kinetics.

Figures \ref{timeseries1} and \ref{hysteresis2} show another example where $%
\sigma _{\infty }$ was increased more gradually as the plate developed. In
this case the plate initially formed at a lower $v_{h}$ compared to the
previous example, and the maximum $v_{r}$ was lower. Again the $v_{h}$-$v_{r}
$ plot shows significant hysteresis, indicative of the SDAK mechanism. Here
we see that $\alpha _{prism}$ increased by a factor of four as the thin
plate developed, while during this same time $v_{h}$ (and thus $\sigma
_{surf})$ declined slightly. Modeling $\sigma _{prism}$ as before yields a
broad dip with a minimum value of $\sigma _{prism}\approx 0.8$ at $t\approx
450$ seconds.

We observed similar plate-on-pedestal growth behaviors at other temperatures
ranging from -12 C to -17 C. The hysteresis in the $v_{h}$-$v_{r}$ plots was
most pronounced at -15 C, and we observed that the minimum $v_{h}$ needed to
produce a transition to plate-like growth was approximately two times higher
at -17 C than at the other temperatures. Overall these observations are
consistent with what we know from the morphology diagram, namely that the
growth of thin, plate-like crystals is most pronounced near -15 C, as shown
in Figure \ref{morphdiagram}.

\begin{figure}[p] % float placement: (h)ere, page (t)op, page (b)ottom, other (p)age
  \centering
  % file name: C:/1KGLaaa/aatempfold/VIGproject/Paper3-EdgeInstability/arXiv/popcomposite2a.jpg
  \includegraphics[width=4.3in, keepaspectratio]{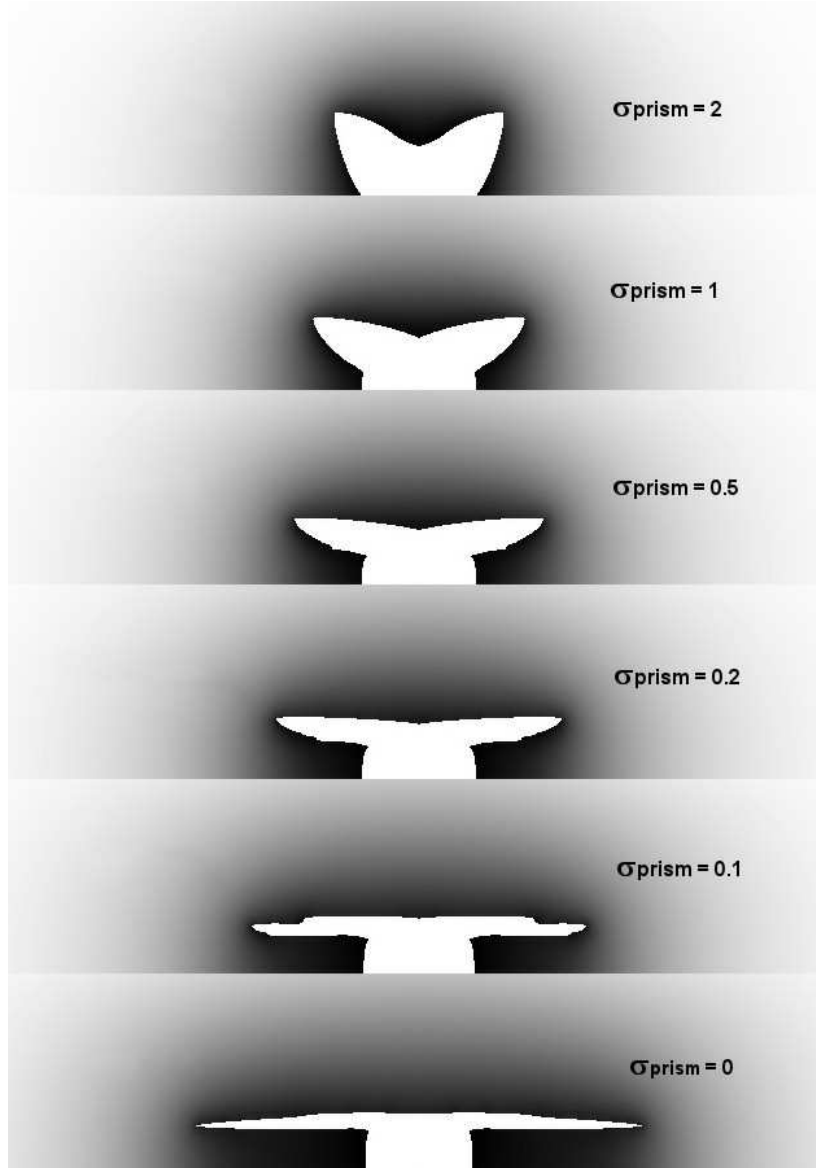}
  \caption{Models of plate-on-pedestal
growth as $\protect\sigma _{prism}$ is varied. Shown are cross-sections of
the cylindrically symmetric crystals after 20 seconds of growth. Brightness
around the crystal is proportional to supersaturation. For each of these
models the initial crystal radius and diameter were both 5 $\protect\mu $m, $%
\protect\sigma _{\infty }=5$ percent, $\protect\alpha _{basal}=\exp (-%
\protect\sigma _{basal,0}/\protect\sigma _{surf}),$ and $\protect\alpha %
_{prism}=\exp (-\protect\sigma _{prism}/\protect\sigma _{surf}).$ Each model
has the same scale, and the $\protect\sigma _{prism}=0$ plate has a final
radius of 26 $\protect\mu $m. The prism growth velocities for the final
crystals shown were $v_{r}=(250,300,450,600,650,900)$ nm/sec, from top to
bottom.}
  \label{popmodels}
\end{figure}

\section{Modeling}

It is also instructive to examine computer models of growing crystals as $%
\sigma _{prism}$ and other parameters are varied, in order to compare with
our observations. To this end we used a cylindrically symmetric cellular
automata model \cite{ca} for which the functions $\alpha _{basal}(\sigma
_{surf})$ and $\alpha _{prism}(\sigma _{surf})$ were inputs in the code. For
these computations we used $\alpha _{basal}(\sigma _{surf})=\exp (-\sigma
_{basal,0}/\sigma _{surf})$ with $\sigma _{basal,0}=2.0$ percent and $\alpha
_{prism}(\sigma _{surf})=\exp (-\sigma _{prism}/\sigma _{surf})$ with a
constant $\sigma _{prism}$ for each model. These models cannot reproduce the
full SDAK growth instability, or the observed hysteresis, because $\sigma
_{prism}$ was assumed to have a constant value independent of the structure
of the crystal. Therefore we used the models mainly to examine the
transition to plate-like growth as a function of $\sigma _{prism}$ and $%
\sigma _{\infty }.$

Figure \ref{popmodels} shows one example of a set of models as $\sigma
_{prism}$ was varied. The figure shows that the models produced thin
plate-like crystals growing from the initial pedestal only when $\sigma
_{prism}$ was quite low, confirming the low $\sigma _{prism}$ values we
inferred from our measurements. From this and additional model calculations
we reached two conclusions:\ 1) both the observed plate-on-pedestal
morphology and the measured $v_{r}$ values were approximately reproduced
with $\sigma _{prism}$ values that were consistent with those inferred from
our experimental data, and 2) the initial growth phase with no thin plate
was reproduced only with larger $\sigma _{prism}$ values, again roughly
consistent with our measured values. Although these constant-$\sigma _{prism}
$ models cannot reproduce the full behavior of our observed crystals, they
nevertheless support our overall conclusions regarding the SDAK mechanism
and the resulting edge-enhancing growth instability.

To produce a full model of the SDAK instability, it would be necessary for $%
\alpha _{prism}$ to vary with the structure of the crystal, as this is an
inherent feature of structure-dependent attachment kinetics. One possibility
would be to let $\alpha _{prism}(\sigma _{surf})=\exp (-\sigma
_{prism}(R)/\sigma _{surf})$ where $R$ is the radius of curvature of the
edge of the crystal plate. The model would assume $\sigma _{prism}(R=\infty
)=\sigma _{prism,0}$, where $\sigma _{prism,0}$ is the measured value for
flat prism facets. At smaller $R,$ $\sigma _{prism}$ would become smaller,
as seen in Figure \ref{sigmacalcs}. Realizing such a dynamical model would
require some guesses for the correct functional form for $\sigma _{prism}(R)$%
, as this function is not determined from our measurements or from
theoretical considerations. Without more constraints on this and other model
inputs, we believe a full dynamical model would be of limited use at this
time, other than as a proof-of-principle that the SDAK instability can
exhibit a hysteresis behavior that approximates what we observed. However,
additional experiments like those described here may yield a useful
measurement of $\sigma _{prism}(R)$, so additional progress in quantitative
dynamical modeling is certainly possible.

\section{Discussion}

To summarize, the data and modeling presented above provide strong evidence
for structure-dependent attachment kinetics and an edge-enhancing growth
instability at -15 C. Our ice crystal growth measurements and modeling
indicate that:

1) For the growth of flat facet surfaces near -15 C, the condensation
coefficients for both the basal and prism facets are well described by the
functional form $\alpha (\sigma )=\exp (-\sigma _{facet,0}/\sigma ),$ where $%
\sigma $ is the supersaturation at the surface and $\sigma _{facet,0}$ is a
constant.

2) For flat facet surfaces we have measured $\sigma _{basal,0}\approx 2.0$
percent and $\sigma _{prism,0}\approx 4.2$ percent at -15 C. Because $\alpha
_{basal,0}<\alpha _{prism,0}$, this suggests that the intrinsic growth
morphology at -15 C is a columnar prism.

3) The measurements suggest that the values of $\sigma _{basal,0}$ and $%
\sigma _{prism,0}$ are independent of background air pressure, at least for
the range 0.01 -- 1 bar. We also expect there would be no intrinsic pressure
dependence in the attachment kinetics from theoretical considerations, since
these pressures are too low to substantially affect the ice surface.

4) While chemical contamination of the ice surface is always present at some
level in experiments, we have found that contamination levels must be quite
high to affect ice growth \cite{contamination}. Furthermore, chemical
contamination typically suppresses the growth of thin plates at -15 C \cite%
{contamination}, suggesting that contamination is not an important factor in
the observations described here.

5) Our observations of plate-on-pedestal growth show a rather abrupt
transition to plate-like growth. Using the basal facet as a
\textquotedblleft witness surface\textquotedblright\ allowed us to estimate $%
\sigma _{prism}$ as described above. During plate growth we found $\sigma
_{prism}\ll \sigma _{prism,0}.$

6) Computer modeling confirms that $\sigma _{prism}\ll \sigma _{prism,0}$ is
necessary for plate-like growth.

7) Hysteresis in the $v_{h}$-$v_{r}$ plots requires that $\alpha
_{prism}(\sigma _{surf})$ is not a simple single-valued function independent
of crystal morphology. Instead, $\alpha _{prism}$ must be large at the edge
of a thin plate and smaller on the edges of thick plates or on flat facet
surfaces.

Together these considerations strongly support the SDAK mechanism we
proposed earlier, and indicate the presence of an edge-enhancing growth
instability. Beginning with a crystal prism containing only large facets, we
initially have $\sigma _{prism}\approx \sigma _{prism,0}$ and $\sigma
_{basal}\approx \sigma _{basal,0}.$ At low pressures the prism and basal
facets remain large and the SDAK mechanism is absent, giving the roughly
isometric growth that has been observed. But at higher pressures, diffusion
causes the top edge of the crystal to grow a bit faster than other parts of
the crystal, so the edge sharpens slightly. This sharper edge leads to a
reduced $\sigma _{prism},$ which causes the edge growth to increase further.
A positive feedback results: as $\sigma _{prism}$ reduces, the edge grows
faster and becomes sharper, thus reducing $\sigma _{prism}$ more, and so on.
This positive feedback is the hallmark of a growth instability, and the end
result is the formation of a thin plate.

An edge-enhancing instability driven by the SDAK mechanism is a relatively
simple physical explanation that nicely explains all the observations. The
existence of a growth instability of this nature may also explain why ice
crystal morphologies change so abruptly with temperature in the morphology
diagram. If the instability tips only slightly in the direction of fast
prism growth, then thin plates will form. But if the same instability tips
toward fast basal growth, then hollow columns (which, like plates, also have
thin edges) become the preferred morphology. Even small changes in the
underlying nature of the instability can result in large changes in the
final crystal morphologies.

We do not yet understand the molecular mechanisms that produce the SDAK
effect, although structure-dependent surface premelting may be playing a
role. We also do not yet understand the temperature dependence of this
instability, and why it is especially prevalent at -15 C. Nevertheless, it
appears that this instability is responsible for the formation of thin
plates at -15 C, which is a dominant feature in the morphology diagram.
Additional observations may yield more insights into the underlying
molecular processes responsible for shaping the growth of snow crystals
under different conditions.

\end{document}